\documentclass[twocolumn]{aastex631}

\usepackage[utf8]{inputenc} 
\usepackage[T1]{fontenc}    
\usepackage{hyperref}
\usepackage{url}
\usepackage{amsfonts}
\usepackage{amsmath}
\usepackage{amssymb}
\usepackage{graphicx}
\graphicspath{{./}{figs/}}
\usepackage[encapsulated]{CJK}
\usepackage{ucs}

\newcommand{\prospector}{\texttt{Prospector}}
\newcommand{\sbip}{\texttt{SBI$^{++}$}}

\shorttitle{SBI$^{++}$: Flexible Likelihood-free Inference}
\shortauthors{Wang et al.}

\begin{document}

\title{SBI$^{++}$: Flexible, Ultra-fast Likelihood-free Inference Customized for Astronomical Applications}

\correspondingauthor{Bingjie Wang}
\email{bwang@psu.edu}

\author[0000-0001-9269-5046]{Bingjie Wang (\begin{CJK*}{UTF8}{gbsn}王冰洁\ignorespacesafterend\end{CJK*})}
\affiliation{Department of Astronomy \& Astrophysics, The Pennsylvania State University, University Park, PA 16802, USA}
\affiliation{Institute for Computational \& Data Sciences, The Pennsylvania State University, University Park, PA 16802, USA}
\affiliation{Institute for Gravitation and the Cosmos, The Pennsylvania State University, University Park, PA 16802, USA}

\author[0000-0001-6755-1315]{Joel Leja}
\affiliation{Department of Astronomy \& Astrophysics, The Pennsylvania State University, University Park, PA 16802, USA}
\affiliation{Institute for Computational \& Data Sciences, The Pennsylvania State University, University Park, PA 16802, USA}
\affiliation{Institute for Gravitation and the Cosmos, The Pennsylvania State University, University Park, PA 16802, USA}

\author[0000-0002-5814-4061]{V. Ashley Villar}
\affiliation{Department of Astronomy \& Astrophysics, The Pennsylvania State University, University Park, PA 16802, USA}
\affiliation{Institute for Computational \& Data Sciences, The Pennsylvania State University, University Park, PA 16802, USA}
\affiliation{Institute for Gravitation and the Cosmos, The Pennsylvania State University, University Park, PA 16802, USA}

\author[0000-0003-2573-9832]{Joshua S. Speagle (\begin{CJK*}{UTF8}{gbsn}沈佳士\ignorespacesafterend\end{CJK*})}
\affiliation{Department of Statistical Sciences, University of Toronto, Toronto, ON M5G 1Z5, Canada}
\affiliation{David A. Dunlap Department of Astronomy \& Astrophysics, University of Toronto, Toronto, ON M5S 3H4, Canada}
\affiliation{Dunlap Institute for Astronomy \& Astrophysics, University of Toronto, Toronto, ON M5S 3H4, Canada}
\affiliation{Data Sciences Institute, University of Toronto, Toronto, ON M5G 1Z5, Canada}

\begin{abstract}

Flagship near-future surveys targeting $10^8-10^9$ galaxies across cosmic time will soon reveal the processes of galaxy assembly in unprecedented resolution. This creates an immediate computational challenge on effective analyses of the full data-set. With simulation-based inference (SBI), it is possible to attain complex posterior distributions with the accuracy of traditional methods but with a $>10^4$ increase in speed. However, it comes with a major limitation. Standard SBI requires the simulated data to have characteristics identical to those of the observed data, which is often violated in astronomical surveys due to inhomogeneous coverage and/or fluctuating sky and telescope conditions. In this work, we present a complete SBI-based methodology, ``\texttt{SBI$^{++}$},'' for treating out-of-distribution measurement errors and missing data. We show that out-of-distribution errors can be approximated by using standard SBI evaluations and that missing data can be marginalized over using SBI evaluations over nearby data realizations in the training set. In addition to the validation set, we apply \texttt{SBI$^{++}$} to galaxies identified in extragalactic images acquired by the James Webb Space Telescope, and show that \texttt{SBI$^{++}$} can infer photometric redshifts at least as accurately as traditional sampling methods---and crucially, better than the original SBI algorithm using training data with a wide range of observational errors. \texttt{SBI$^{++}$} retains the fast inference speed of $\sim$~1 sec for objects in the observational training set distribution, and additionally permits parameter inference outside of the trained noise and data at $\sim$~1 min per object. This expanded regime has broad implications for future applications to astronomical surveys~\footnote{Code and a Jupyter tutorial are made publicly available at \url{https://github.com/wangbingjie/sbi_pp}.}.

\end{abstract}


\section{Introduction}

Spectral energy distribution (SED) fitting is the primary basis on which observations of galaxies across cosmic time are grounded to theories of galaxy formation and evolution. The traditional Bayesian approach involves the following components: the generation of model galaxy SEDs using stellar population synthesis (SPS) given a set of input physical parameters, expectation of the probability of various solutions encoded as priors, and a sampler. The most common approach in astronomy today is combining SPS models (e.g., \citealt{Bruzual2003,Conroy2010}) with Markov chain Monte Carlo (MCMC; \citealt{Goodman2010}) or nested sampling (NS; \citealt{Skilling2004}) to attain posterior distributions in a Bayesian inference framework. This family of codes includes, for example, \texttt{BAGPIPES} \citep{Carnall2018}, \texttt{BEAGLE} \citep{Chevallard2016}, and \texttt{Prospector} \citep{Johnson2021}. Since galaxies are inherently complex systems, such an approach generally requires the generation of $\sim$~1-2 million models for each object. At $\sim$~0.05s per model, this translates to $\sim$~100 billion CPU-hours to fit the billions of galaxies expected to be observed by the Vera C. Rubin Observatory \citep{2019ApJ...873..111I}. Within a few years, the Rubin Observatory will observe its first light, which creates an urgent need to address this computational challenge.

A number of techniques have been considered in the literature to increase the speed of inference. First, the generation of an SPS model can be accelerated by using a differentiable SPS code or an artificial neural network emulator. The former generates exact SPS with a boost in speed enabled by specific code libraries (e.g., \citealt{Hearin2021}); while the latter uses a quick-to-evaluate neural network that approximates SPS (e.g., \citealt{Alsing2020}). Second, the sampling time can be decreased by switching to a gradient-based sampler \citep{Duane1987,Hoffman2011}, or combining the sampler with a normalizing flow \citep{Karamanis2022,2023JOSS....8.5021W}. Currently, the only technique that is mature enough to have the flexibility and the capability on par with the traditional method is a neural net emulator + MCMC/NS at a cost of $\gtrsim 15$ mins per fit \citet{Alsing2020,Mathews2023}. While fast and reliable, this still is not quick enough: 15 mins per fit requires a billion CPU hours to fit all galaxies expected to be observed by the Rubin Observatory.

It is not obvious how to accelerate model generation further using current techniques; the next step forward therefore would most likely need to bypass the likelihood calculation entirely. Simulation-based inference (SBI), which uses normalizing flows to learn posterior densities directly, is the ideal candidate. Several recent works have already adopted such neural density estimators to analyze astrophysical data (e.g., \citealt{Alsing2019,Green2020,Dax2021,Zhang2021,DIGS2022,Leja2022,Ting2022}; see \citealt{Cranmer2019} for a recent review). However, the drastically increased evaluation speed of SBI is accompanied by an inflexibility: the data to be modeled must have identical properties as the training data, including nearly identical noise properties, free parameters, exact priors, and so forth. This limits the applicability of SBI to astronomical surveys because the implied assumptions---stable noise properties and complete input data---are often violated in real astronomical data. Varying telescope and sky conditions, and heterogeneous data coverage are common, in particular when combining data sets across multiple surveys for greater coverage in wavelengths or area.

The unusability of SBI on incomplete data has further ramifications for JWST surveys. A filter on the JWST/NIRCam instrument is defined by two different transmission curves depending on whether the measurements fall on module A or B of the camera. A mixture of data taken with module A and B filters for any given object is largely unavoidable, particularly if combining data across different surveys, meaning that SBI applied to JWST data will almost always have a significant number of ``missing'' bands corresponding to the unused filter curve in the other module. Expanding standard SBI to treat missing data is thus necessary for achieving the maximum science return from JWST observations.

In this Letter, we cover two topics. We first show that standard SBI can accurately infer joint posterior distributions of photometric redshifts and key stellar populations metrics. This is a more challenging case compared to \citet{Hahn2022}, where redshift is a fixed parameter. We opt for this approach given the broad interest in deriving accurate photometric redshifts and also to ensure that the often non-Gaussian uncertainties associated with redshifts can be properly propagated into the physical parameters describing the galaxy populations. Then, building on our baseline SBI, we present \sbip---an SBI-based methodology to deal with large noise and missing photometric bands. It requires no additional model training, but expands the applicability of SBI to astronomical data.

This Letter is structured as follows. Section~\ref{sec:sbi} outlines the theoretical framework of standard SBI as well as \sbip. Section~\ref{sec:baseline} presents our baseline SBI model, and assesses its performance using a simulated data set. Section~\ref{sec:sbip} describes the new methodology \sbip, which includes details of its implementation, evaluation speed, and accuracy. Section~\ref{sec:jwst} deploys \sbip\ on a recently acquired JWST data set as a further demonstration of its applicability. Section~\ref{sec:concl} discusses the key findings, and concludes with implications for future works. All magnitudes in this Letter are expressed in asinh AB magnitudes with a floor of 35 magnitude \citep{Lupton1999}. Where applicable, we adopt the best-fit cosmological parameters from the 9 yr results from the Wilkinson Microwave Anisotropy Probe mission: $H_{0}=69.32$ ${\rm km \,s^{-1} \,Mpc^{-1}}$, $\Omega_{M}=0.2865$, and $\Omega_{\Lambda}=0.7135$ \citep{Bennett2013}.

\section{Mathematical Framework\label{sec:sbi}}

Bayes' theorem states that
\begin{equation}
	P(\theta|x) = \frac{P(x|\theta)P(\theta)}{P(x)},
\end{equation}
where $x$ is a vector containing observations and $\theta$ is a vector containing model parameters. The goal of Bayesian SED modeling is to infer the posterior $P(\theta|x)$ by taking advantage of our prior knowledge $P(\theta)$ and mapping out the likelihood surface $P(x|\theta)$. $P(x)$ is the (often neglected) model evidence, which is a normalizing factor useful primarily for comparison between different models.

Traditional methods for attaining $P(\theta|x)$ involves advanced sampling methods, the most popular implementations of which in astronomy are MCMC \citep{Foreman-Mackey2013} and NS \citep{Feroz2009,Handley2015,Speagle2020,Buchner2021}. The fundamental difference between MCMC and NS is that the former tries to sample from $P(\theta)$ directly, while the latter samples from slices of $P(\theta)$ that have simpler distributions. NS tends to perform better at multi-modal distributions and possesses well-defined stopping criteria. Given that posteriors are often multi-modal in galaxy SED-fitting (e.g., redshift, or the age-dust-metallicity degeneracy), we choose the NS code \texttt{dynesty} \citep{Speagle2020} as the point of reference to which we compare the performance of our baseline SBI model and \sbip.

\subsection{Standard Simulation-based Inference\label{subsec:overview}}

In contrast to the traditional methods mentioned above, SBI bypasses the likelihood framework entirely. Given a large dataset of $n$ parameter-data pairs $\{\theta_i, x_i\}_{i=1}^{n}$, SBI learns the joint density directly with hyperparameters $\phi$:
\begin{equation}
    \{\theta_i, x_i\} \hookrightarrow P_{\phi}(\theta, x) \approx P(\theta, x),
\end{equation}
where we have explicitly included the $P_\phi(\cdot)$ notation to emphasize that this is an approximation to the true density $P(\cdot)$.

SBI based on density estimation is amortized: once a neural density estimator is trained, the computationally expensive steps involving millions of model evaluations do not have to be repeated for new observations. This is particularly desirable for our application in SED modeling, given that individual model calls are very computationally expensive.

However, a major impediment to applying SBI to full astronomical surveys is its stringent requirement that the training set has identical characteristics to the observed data. Standard SBI performs very well given noise identical to that of the training set \citep{Hahn2022}. Noise is typically included by injecting the errors into the training set, and then conditioning on them. This means that we can generate $n$ $\{\theta_i, x_i, \sigma_i\}_{i=1}^{n}$ pairs, which are then used to learn the joint density using the same strategy as above via
\begin{equation}
    P(\theta|x,\sigma) \approx P_\phi(\theta|x,\sigma) = \frac{P_\phi(\theta, x,\sigma)}{P_\phi(x,\sigma)} \propto P_\phi(\theta, x, \sigma).
\end{equation}

While we expect most observational noise can be captured by a carefully chosen noise model, in practice, even generous training sets will not cover the wide range of observed noise distributions. To solve this problem, we propose to use baseline SBI to marginalize over possible noise values via simple Monte Carlo (MC) integration as outlined in the following section.

\subsection{Out-of-distribution Measurement Errors\label{subsec:math_mcnoise}}

Assume an observed value, $x_i$, has measurement uncertainties, $\sigma_i$, that lie outside of the training set. Using Bayes' Theorem and refactoring a few terms, this means we need to solve the following integral:
\begin{equation}
    P_\phi(\theta|x,\sigma) \propto \int_{\Omega(x^*)} P_\phi(\theta, x^*) P(x | x^*, \sigma) \, {\rm d} x^*,
\end{equation}
where $x^*$ is the true value with no measurement uncertainties, $\Omega(x^*)$ is the entire (finite) domain of possible $x^*$ values, $P_\phi(\theta, x*)$ is the probability density function (PDF) derived from $\{\theta_i, x_i^*\}$ pairs, and $P(x | x^*, \sigma)$ is the possibly unknown and/or analytically intractable PDF associated with the noise process. Given a sample of $m$ simulated values $\{x^*_1,\dots,x^*_m\}$ from this PDF, we can construct an MC approximation of the integral as
\begin{equation} \label{eq:sbi}
    \int_{\Omega(x^*)} P_\phi(\theta, x^*) P(x | x^*, \sigma) \, {\rm d} x^* \approx \frac{1}{m} \sum_{j=1}^{m} P_\phi(\theta, x^*_j);
\end{equation}
i.e., a sum over repeated evaluations of SBI.

\subsection{Missing Data\label{subsec:math_missdata}}

Standard SBI does not allow missing data. One can also think of the missing bands as data where $\sigma_i \rightarrow \infty$, which means that we can assume $P(x|x^*,\sigma) \approx C$ over the entire (finite) domain $\Omega(x^*)$. If we separate out $x = \{x_{\rm o}, x_{\rm m} \}$ and $\sigma = \{ \sigma_{\rm o}, \sigma_{\rm m}\}$ into observed $\{ x_{\rm o}, \sigma_{\rm o}\}$ and missing $\{x_{\rm m}, \sigma_{\rm m}=\infty\}$ values, we can define the integral we need to solve more explicitly as
\begin{equation}
    P_\phi(\theta|x,\sigma) \propto \int_{\Omega(x^*)} P_\phi(\theta, x^*) P(x_{\rm o} | x^*_{\rm o}, \sigma_{\rm o}) \, {\rm d} x^*_{\rm o} {\rm d} x^*_{\rm m},
\end{equation}
where $x^* = \{x^*_{\rm o}, x^*_{\rm m} \}$ is again the true value that can be broken up into observed $x^*_{\rm o}$ and missing $x^*_{\rm m}$ components.

In practice, a small sub-domain contributes to most of the integral, which simplifies the evaluation. As a result, we approximate missing data by using kernel density estimation (KDE) based on the distribution of nearby neighbors (NNs) in the observed bands within our training set to define a local density function $Q(x^*_{{\rm m}} | x^*_{{\rm o}})$ for the missing bands given the observed ones (see Section \ref{subsec:sbip_imp} for additional details). Assuming the true distribution $P(x^*_{{\rm m},j} | x^*_{{\rm o}, j}) \approx Q(x^*_{{\rm m}} | x^*_{{\rm o}})$, we can use our NN approximation to simulate values of $x^*_{{\rm m},j}$ given a simulated value of $x^*_{{\rm o},j}$ and arrive at the updated approximation
\begin{equation}
P_\phi(\theta|x,\sigma) \approx \frac{1}{m} \sum_{j=1}^{m} \frac{P_\phi(\theta, x^*_j)}{Q(x^*_{{\rm m},j} | x^*_{{\rm o}, j})},
\end{equation}
which is just a re-weighted version of our approximation from Equation \eqref{eq:sbi}.
It should be noted that the importance weighting performed here (i.e., proposing from $Q(x^*_{{\rm m}} | x^*_{{\rm o}})$ but targeting $P(x^*_{{\rm m},j} | x^*_{{\rm o}, j})$) removes the impact that the exact data neighborhood has on the inference. Our final estimate then is mostly sensitive to the overall coverage of the NN-based proposal.

In essence, the two methods introduced above aim to construct the input data in a way that is appropriate for the baseline SBI, and approximate the true posterior distribution by averaging over MC samples drawn from the observed distribution. Therefore, as long as we can write down an approximation in the data space, the MC technique is flexible enough to deal with other systematics related to the data. However, it does not address modeling systematics or unknown unknowns. Modeling assumptions are integrated in the training phase, so a model mismatch problem, for instance, would require a reconsideration of the training of the flow. As for the unknown unknowns, if they could be marginalized over (e.g., by including an extra scatter term), then one would need to modify the architecture appropriately or incorporate it into the MC integration procedure.

\section{Baseline SBI\label{sec:baseline}}

\begin{figure*}
\gridline{
\fig{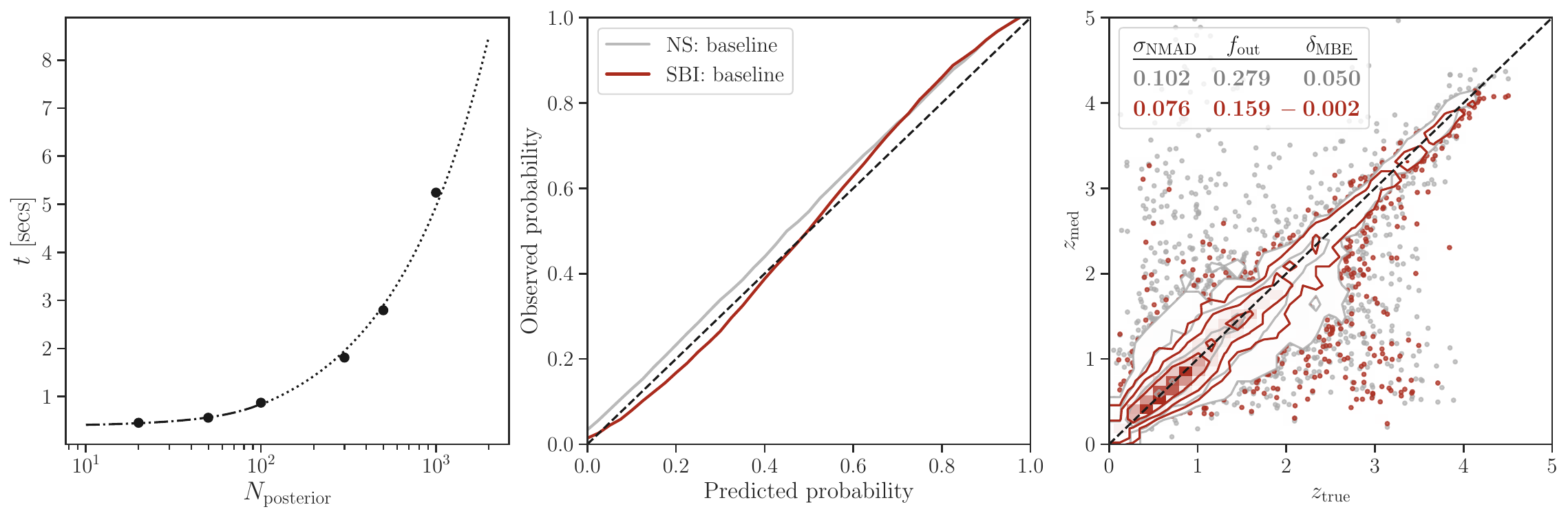}{0.99\textwidth}{}
}
\caption{Here we show the performance of the baseline SBI. (a) Runtime as a function of the number of posterior samples drawn in each SBI call. The dotted-dashed line shows the exponential fit to the data, whereas the dotted line shows the power-law fit in Equation~\ref{eq:fit_baseline}. (b) Predictive-probability plot comparing the cumulative distributions of redshifts. The nested sampling result is shown in gray, while the SBI result is shown in red. Both lines follow along the diagonal, meaning both models are well calibrated. (c) Contours showing the median of the redshift posteriors as a function of true redshifts. Interestingly, the baseline SBI infers redshifts more accurately than NS. This suggests that the standard SBI performs exceptionally well on data closely following the training set distribution. \label{fig:baseline}}
\end{figure*}

Having laid out the mathematical foundation, we now describe the training of our baseline SBI model, which will also be used in \sbip. The novelty of the baseline necessitates a dedicated assessment of its performance, including evaluation speed and accuracy. We do so on simulated galaxies in Sections~\ref{subsec:baseline_speed} and \ref{subsec:baseline_acc}, respectively, before presenting \sbip.

\subsection{Training the Neural Density Estimator\label{subsec:exp}}

The training set consists of $\sim$~2 million sets of model SEDs and corresponding galaxy properties used to generate them. The validation set contains around 5,000 held-out examples, drawn from the same distribution as the training set. The size of the validation set is limited by the computational time required to estimate posterior quantities with NS (it takes $\gtrsim$ 15 mins per fit with the aid of a neural net emulator mimicking SPS models; \citealt{Alsing2020,Mathews2023}). We simulate mock photometry composed of 7 HST and 5 JWST bands spanning $0.4 - 5 \mu$m in the observed frame. We adopt the \prospector-$\beta$ galaxy physical model \citep{Wang2023}, which builds on \prospector-$\alpha$ \citep{Leja2017} but with the addition of three joint priors encoding empirical constraints of redshifts, masses, and star formation histories in the galaxy population. In all, the model consists of 18 free parameters describing the contribution of stars, AGNs, gas and dust to galaxy SEDs \citep{Charlot2000,2006MNRAS.371..703S,Noll2009,Conroy2010,Choi2016,Dotter2016}. The star formation history is described non-parametrically by mass formed in 7 logarithmically-spaced time bins, and assumes a continuity prior to ensure smooth transitions between time bins \citep{Leja2019}. The noise is propagated into the training set by assuming a Gaussian noise distribution in magnitude space.

We then construct our model using the Masked Autoregressive Flow \citep{2017arXiv170507057P} implementation in the \verb|sbi| Python package \citep{2019arXiv190507488G,tejero-cantero2020}. The model has 15 blocks, each with 2 hidden layers and 500 hidden units. Training our model takes $\lesssim$ two days on a single NVIDIA Tesla K80 GPU.

\subsection{Evaluation Speed\label{subsec:baseline_speed}}

We plot the runtime as a function of the number of posterior samples, $N$, in Figure~\ref{fig:baseline}(a). The runtime is generally on the scale of seconds, and can be fitted by a combination of exponential and power-law functions as
\begin{equation}
\begin{split}
	t = & \, 0.379 \, {\rm exp} \left( 0.008N \right), \, {\rm for} \, N \leq 100; \\ {\rm log_{10}} \left( t \right) = & \, 0.775 \, {\rm log_{10}}\left( N \right) - 1.630, \, {\rm for} \, N > 100.
\end{split}
\label{eq:fit_baseline}
\end{equation} 
The runtimes are all evaluated on a standard 2020 Macbook Pro laptop using CPUs.

\subsection{Assessing the Accuracy of Recovered Posteriors\label{subsec:baseline_acc}}

The recovery fidelity of the most important and challenging physical parameter, redshift, is shown in Figures~\ref{fig:baseline}(b)-(c). The predictive-probability (p-p) plot shows the predicted cumulative distribution against the observed cumulative distribution. The fact that both NS and SBI results follow along the diagonal indicates that both models are well-calibrated. We additionally show the posterior median as a function of truth. The scatter in residuals is quantified using the normalized median absolute deviation (NMAD). It is commonly used in photometric redshift studies (e.g., \citealt{Dahlen2013,Skelton2014}) as it is less sensitive to outliers than standard indicators such as the root mean square. NMAD is defined as
\begin{equation}
	\sigma_{\rm NMAD} = 1.48 \times {\rm median |\Delta \theta|},
\end{equation}
where $\Delta \theta$ is the difference between two data sets. Here, we replace $\Delta \theta$ with $\Delta z = (z_{\rm phot} - z_{\rm spec}) / (1+z_{\rm spec})$ for consistency with the convention in photometric studies. We also quantify an outlier fraction, $f_{\rm out}$, in which we define a catastrophic outlier as one with $|\Delta z| > 0.15$, and bias calculated using the mean bias error (MBE) as
\begin{equation}
	\delta_{\rm MBE} = \frac{1}{n} \sum_{i=1}^{n} \Delta \theta.
\end{equation}

Surprisingly, the baseline SBI is noticeably better at getting the correct redshifts, which is reminiscent of an earlier finding on comparing a customized MCMC application and a brute-force grid search \citep{Speagle2016}. Here, since the test is done on simulations, where we inject Gaussian noise, the likelihood is known perfectly. The difference in the results is thus purely due to the different sampling methods.
We find that the larger outlier fraction and bias in NS results are mainly driven by the cases where NS incorrectly settles in a high-$z$ solution, whereas SBI places more posterior mass at the true low-$z$ mode. The reverse does not hold true; that is, NS and SBI are equally likely to underestimate redshifts. This suggests that SBI is highly influenced by priors since our informative prior disfavors high-$z$ solutions. Seen in another light, SBI performs exceptionally well on data that closely follow the training set distribution.

\section{\sbip: SBI for astronomical applications\label{sec:sbip}}

\begin{figure*}
  \centering
  \includegraphics[width=0.99\textwidth]{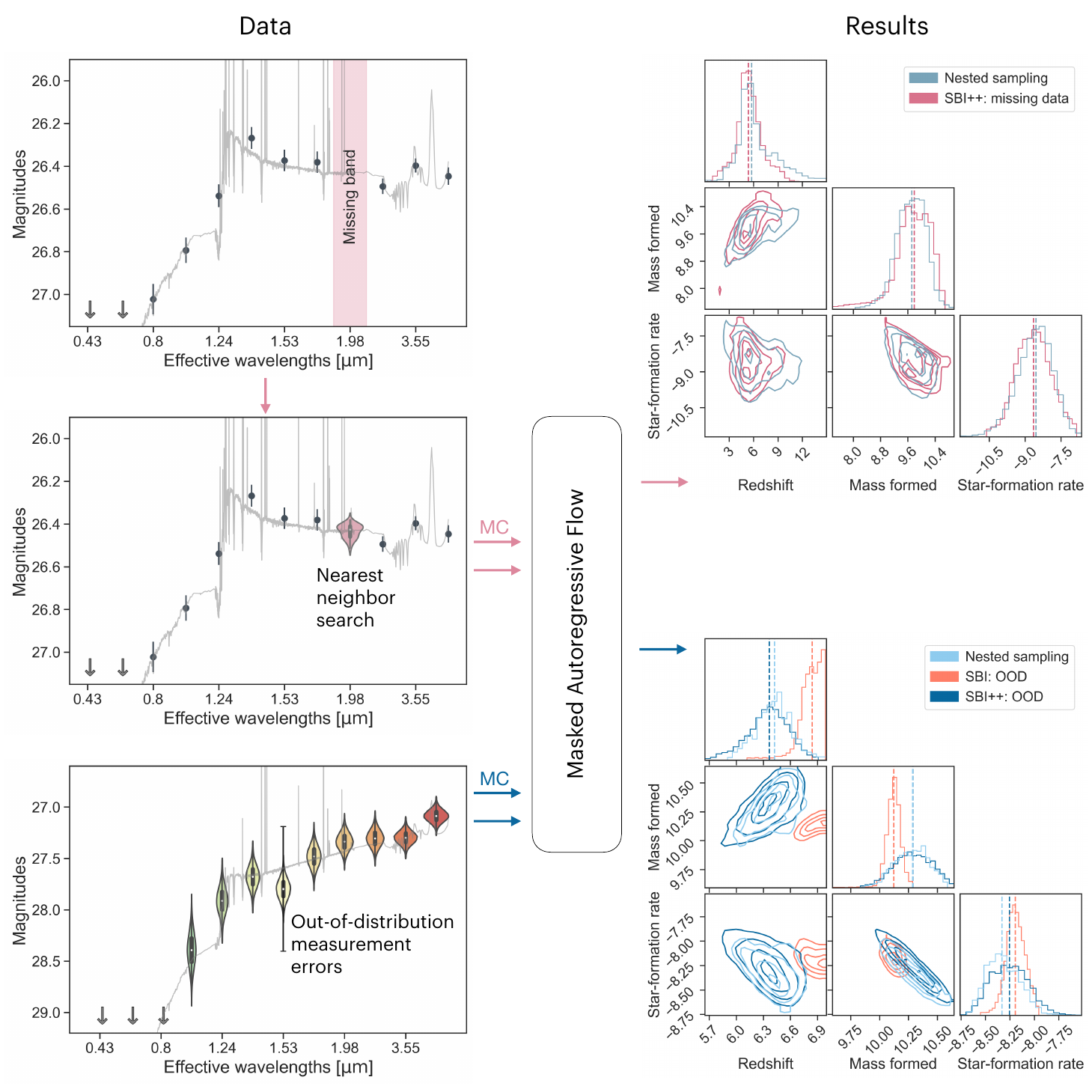}
  \caption{Schematic diagram showing our algorithm, \sbip, for dealing with missing data and OOD measurement errors. First, the SED on the top left has one band missing, rendering it inaccessible to our baseline SBI. Its approximate solution (middle left panel) is found by NN search along with MC integration. The resulting posteriors (top right) show good agreement with NS. Second, the violin plot on the bottom left shows one of our simulated SEDs, with Gaussian noise added to the true underlying SED. Given OOD uncertainties (black error bars), we marginalize over possible noise by MC integration. The bottom right corner plot shows the different posteriors from NS, the naive usage of baseline SBI, and \sbip. Notably, our method performs similarly to the traditional NS method and markedly better than the naive SBI. 
 \label{fig:sche}}
\end{figure*}

Here, we present \sbip, which uses tools native to the standard SBI, but deals with out-of-distribution (OOD) measurement errors and missing data. A schematic representation is shown as Figure~\ref{fig:sche}.

\subsection{Implementing \sbip\label{subsec:sbip_imp}}

\subsubsection{Out-of-distribution Measurement Errors}

As mentioned earlier, we model noise as a Gaussian in uncertainty at fixed magnitude. This means that, for a given observation, we can calculate the expected $1\sigma$ width of the distribution of uncertainties, $\sigma_\star$, from our toy noise model. This expected width is then compared to the observed uncertainty, $\sigma_{\rm o}$. If the difference is greater than a certain threshold, the measurement is marked as OOD. In this work, we define OOD with respect to the toy noise model as any data that shows $(\sigma_{\rm o}-\sigma_\star)/\sigma_\star > 3$. This definition can be generalized to non-Gaussian noise properties by using, for instance, quantiles instead.

After identifying an OOD measurement, we create a set of simulated photometry drawing from a Gaussian distribution with a mean of the observed value and a standard deviation of the observed uncertainty. Note that this can easily be generalized to non-Gaussian noise properties, so long as the user can specify the distribution in test time. Each simulated photometry is assigned an uncertainty of the mean in the noise model at its magnitude. These measurements are passed through the baseline SBI model to produce posteriors; subsequently averaging over all the ``noisy'' posterior samples provides the final parameter estimations.

We note that the MC process may generate noisy data which lie outside of one's model space entirely, causing dangerous extrapolation within the SBI machinery. To mitigate this problem, we truncate a given Gaussian noise distribution to be within a reasonable range that is estimated from an ensemble of similar SEDs in the training set. Specifically, this range is determined based on NNs chosen in magnitude space satisfying a reduced $\chi^2 \leq 5$ ($\chi^2_{\rm red} = \chi^2/n_{\rm bands}$). In the occasional case where there is an insufficient number of NNs ($n \leq 10$), we increase the cut on $\chi^2_{\rm red}$ in increments of 5. This effectively removes the improbable cases where the MC sample representing the noisy band differs from the neighboring bands by several magnitudes.

There are two numbers that need to be determined empirically. First, the threshold over which the trained SBI becomes unreliable depends on the noise model. In general, while one can increase the usability of SBI by increasing noise spread at fixed magnitude, this comes at a significant cost in accuracy. \sbip\ retains the accuracy, although at the expense of an increase in evaluation time as we shall see in Section~\ref{subsec:res_mcnoise}. Therefore, the construction of the noise model, and subsequently, the condition triggering \sbip, are to be decided by balancing the trade-off between accuracy and overall runtime.

Second, the total number of posterior samples required is subject to the complexity of the posterior distribution. This number is, in fact, a product of the number of MC samples and the number of posterior sample drawn for each band of photometry. We find $\gtrsim 100$ MC samples and $\gtrsim 50$ posterior samples are generally more than sufficient for our purpose. In general, the required number of samples will scale with the particulars of the problem, e.g., the complexity of likelihood space and the dimensionality of the data or model.

\subsubsection{Missing Data}

As for the missing data case, we start by finding all SEDs in the training set whose reduced-$\chi^2$ calculated with respect to the observed SED are $\leq$ 5. Then, we construct a KDE from those NNs, weighted by the inverse of their Euclidean distances to the observed SED, for each of the missing bands. Finally, we draw random samples from the KDE, and pass them to the baseline SBI and average over the posteriors.
It is worth noting that in the presence of a sparsely-sample training set, and/or unreasonable priors, the possible non-uniform distribution of points in the neighborhood of a given data measurement may bias the posteriors. 
In our case, the priors are smooth, and so they have little effect on local scales in general. The definition of locality, however, is an empirical question, and varies based on the data. Under the normal circumstance of sufficient observed information, the local density is small, and hence the change in priors over this space is unimportant. Conversely, if the number of missing data points significantly exceeds the number of observed data points, then the local density approaches the limit of the whole prior space, and the priors certainly impact the posteriors.
A recommended practice is to compare a subset of the reconstructed missing bands to the simulated SEDs from the forward model in order to confirm that the local density distribution is in general a reasonable approximation.

\subsection{Results\label{subsec:sbip_res}}

\subsubsection{Evaluation Speed}

\begin{figure*}
\gridline{
\fig{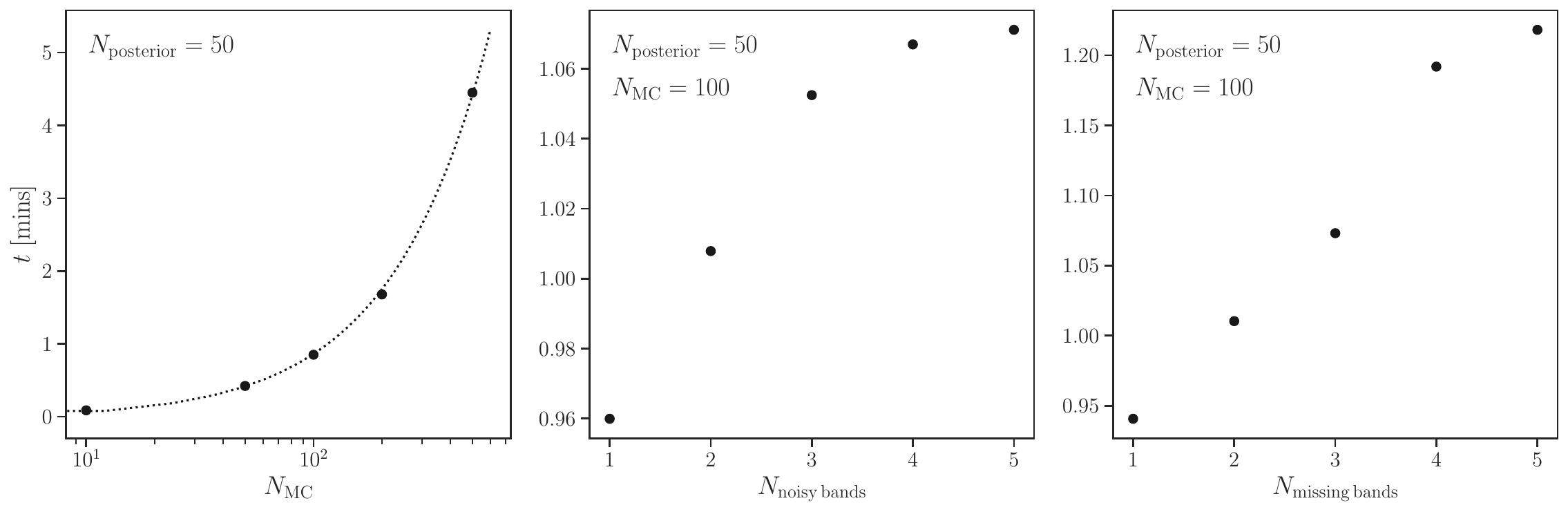}{0.99\textwidth}{}
}
\caption{Here we show the runtime of \sbip\ under different settings. 
(a) Runtime as a function of the number of MC samples drawn in \sbip, and each MC sample contains 50 posterior samples. The dashed line indicates the power-law fit of Equation~\ref{eq:fit_sbip}. (b) Runtime as a function of the number of noisy bands. We draw 100 MC samples, where each MC sample contain 50 posterior samples. (c) Runtime as a function of the number of missing bands. The deviations from a power law shown in these two plots is likely due to the NN search. We additionally note that the runtime of \sbip\ does not depend on the exact band that is noisy/missing; that is, it takes \sbip\ roughly the same time to approximate a blue band as a redder band. We caution, however, these figures only serve as illustrations for the dependence of runtime, but not a guide for determining these parameters. The choice of these parameters depends on the data and the science question. \label{fig:time}}
\end{figure*}

Using \sbip\ to cover OOD noise and missing data takes $\sim$~1 min per fit due to the multiple MC draws required. Additional details on the execution time of different settings are supplied in Figure~\ref{fig:time}. The first panel shows the runtime as a function of number of MC samples, $N_{\rm MC}$, when 50 posterior samples are drawn per MC sample. The following power-law fit illustrates how this runtime (in seconds) scales with increasing $N_{\rm MC}$:
\begin{equation}
	t = 0.534 N_{\rm MC} - 1.685, \, {\rm for} \, N_{\rm MC} \gtrsim 10.
	\label{eq:fit_sbip}
\end{equation}

The other two panels show the execution time as a function of the number of noisy or missing bands, respectively. While slower than baseline SBI, \sbip\ is still $\sim 500\times$ faster than traditional methods and requires no additional model evaluations or neural net training beyond what is needed for baseline SBI. We additionally note that the number and particular configuration of the noisy/missing bands has negligible impact on the execution time.

\subsubsection{Assessing the Accuracy of SBI++ in the Out-of-distribution Noise Approximation\label{subsec:res_mcnoise}}

\begin{figure*}
  \centering
  \includegraphics[width=1\textwidth]{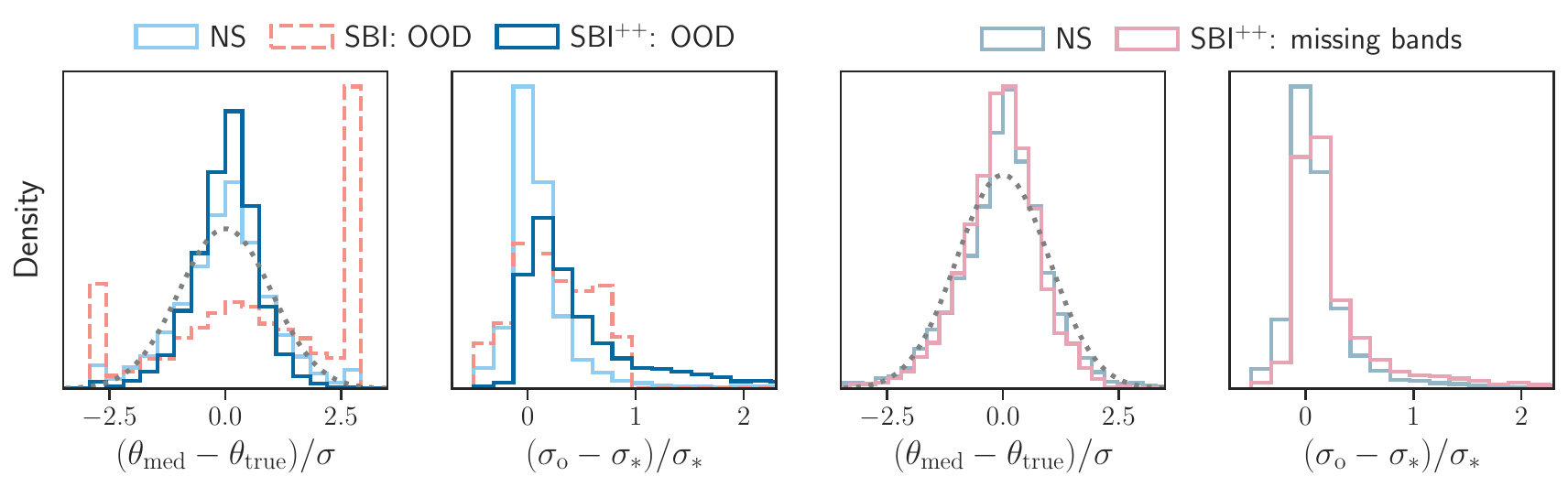}
  \caption{The two panels on the left illustrate the changes in SBI/NS redshift posteriors estimated from the noisy photometry with respect to those from the unperturbed photometry. Similarly, the two panels on the right describe the changes in SBI/NS posteriors estimated from incomplete photometric data with respect to those from the complete data. Unit Gaussians are over-plotted as gray dotted lines to guide the eye. It is evident that \sbip\ recovers the parameters with accuracy comparable to that of a standard inference methodology like NS. We also show results from improperly using the baseline SBI when the noise is OOD to demonstrate the necessity of applying our method. The $\delta_{\sigma,{\rm OOD}}<0$ group manifested in the second orange histogram shows naive SBI finds the wrong solution but with high confidence.
  \label{fig:hist}}
\end{figure*}

To assess the accuracy of \sbip\ when treating OOD noise, we inflate the noise by $5 \sigma$ in a random band for objects in the validation set, and compare results from \sbip\ (OOD) and NS. We start by assessing the absolute accuracy using the same statistics as in Section~\ref{subsec:baseline_acc}. 
For NS and \sbip\ (OOD) results respectively, the NMADs are 0.11 and 0.09, the outlier fractions are 0.42 and 0.36, and the MBEs are 0.05 and 0.04. The agreement here is excellent, although we need to reconcile with the previously seen better performance of the baseline SBI on the validation set in Section~\ref{subsec:baseline_acc}. We do so in Section~\ref{sec:concl}. In contrast, improperly using the baseline SBI when the noise is OOD produces significantly less accurate results as expected. In this case, the NMAD is 0.35, the outlier fraction is 0.69, and the MBE is 0.35.

We further evaluate the results by calculating the shifts in medians and standard deviations of the posteriors generated from SBI (baseline) and \sbip\ (OOD). We also compare to the shifts in NS posteriors. The shift in medians is quantified as
 \begin{equation}
 	\delta_{\rm med} = (\theta_{\rm med} - \theta_{\rm true}) / \sigma,
 	\label{eq:delta_med}
 \end{equation}
where $\theta$ is the parameter of interest, and $\sigma$ is the $(84^{\rm th} - 16^{\rm th})/2$ quantile width in the posterior distribution. The shift in standard deviations is estimated as
 \begin{equation}
 	\delta_{\sigma} = (\sigma_{\rm o} - \sigma_{\ast}) / \sigma_{\ast},
 	\label{eq:delta_sig}
 \end{equation}
where $\sigma_{\rm o}$ is the standard deviation of posteriors predicted from the noisy photometry, and $\sigma_{\ast}$ is that from the original photometry. This approach is chosen over tests such as KL-divergence given the known difficulty of statistical tests in multivariate settings \citep{Kullback1968}. Here we simply evaluate whether the shifts seen in the \sbip\ are within reasonable expectation.

The redshift results are shown in Figure~\ref{fig:hist}. Other parameters exhibit similar trends. The offsets of the posterior median from the truth seen in \sbip\ and NS show excellent agreement, as evident from the first panel. Results from the improper usage of baseline SBI again are much worse. The change in the uncertainties $\delta_{\sigma}$ shown in the second panel, however, exhibits more complex features. We find that this is mostly due to the multi-modality of the posterior distribution, and we provide a detailed discussion in Section~\ref{sec:concl}.

\subsubsection{Assessing the Accuracy of Nearest Neighbor Search for Missing Bands\label{subsec:res_missdata}}

We assess the accuracy of \sbip\ (missing data) by randomly masking a band in the validation set. For NS and \sbip\ results, the NMADs are 0.11 and 0.09, the outlier fractions are 0.43 and 0.34, and the MBEs are 0.05 and 0.01, respectively. The performances again are similar.

We likewise compare the shift in medians and standard deviations of the posteriors. The results, also shown in Figure~\ref{fig:hist}, demonstrate that \sbip\ performs comparably to NS.

\section{Applying \sbip\ to JWST Observations\label{sec:jwst}}

\begin{figure*}
\gridline{
\fig{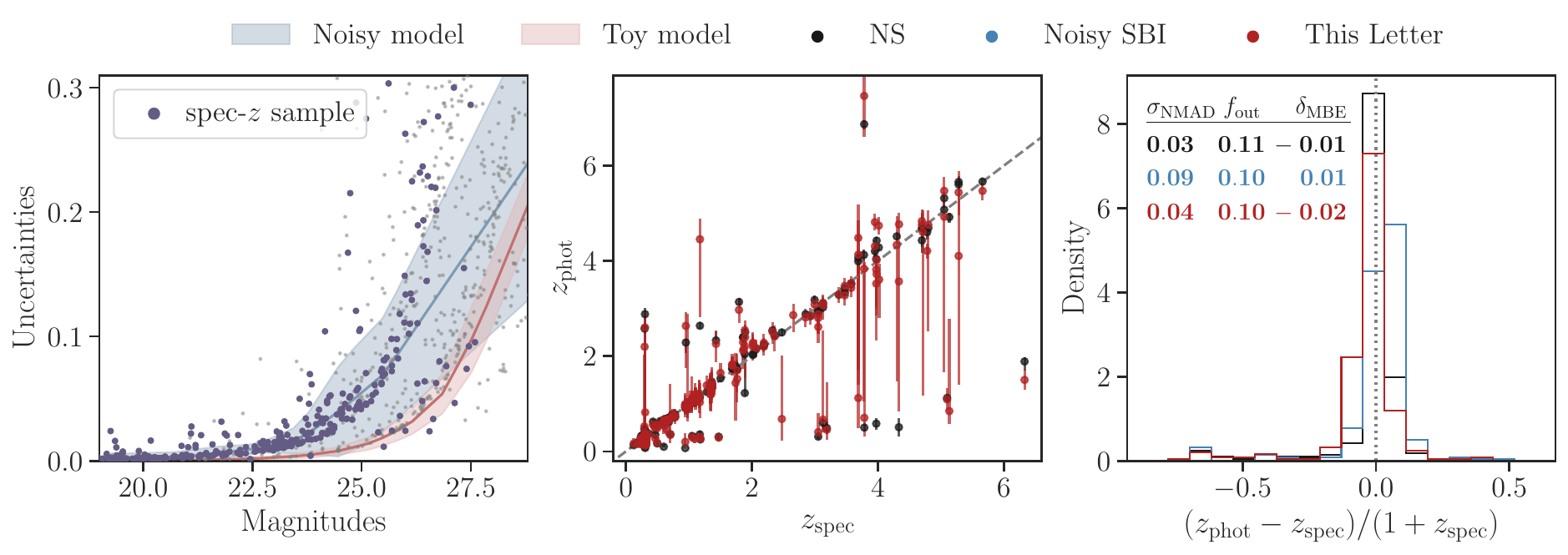}{0.99\textwidth}{}
}
\caption{Here we apply \sbip\ to JWST observations and compare its photometric redshift recovery to NS. (a) The scatter plot shows the measurement uncertainties as a function of magnitudes in the F444W band. The full sample is shown in gray, with the spectroscopic subsample highlighted in purple. The red shades indicate our toy noise model, while the blue shades indicates a ``wider'' noise model that encapsulates more data points. (b) Inferred photometric redshifts are plotted against the spectroscopic redshifts, which we take to be the truths. NS results are shown in black, while \sbip\ results are shown in red. Interestingly, \sbip\ appears to produce more realistic uncertainties. (c) The histograms show the residuals in the inferred redshifts. NS and \sbip\ exhibit comparable performance. However, the noisy SBI model, shown in blue, results in a scatter about twice as large as those of the other two methods.\label{fig:jwst}}
\end{figure*}

In addition to the validation set, we examine the performance of \sbip\ on real extragalactic observations recently acquired by JWST in the Abell 2744 galaxy cluster field (DD 2756; PI Chen). This is a more stringent test as it implicitly includes potential real-world effects such as model mismatches, incorrectly estimated noise, or similar practical issues. Our data consists of 5 JWST bands, and 7 archival HST bands. The data reduction follows \citet{Weaver2023}. 

As mentioned in Section~\ref{sec:baseline}, our training set is generated from the Prospector-$\beta$ prior distributions developed in \citet{Wang2023}. These priors are based on well-established observations of galaxy evolution, and hence are helpful in determining the more probable solutions. However, informative priors also lead to an undesirable consequence of sparse training set coverage in parts of the parameter space. We mitigate this problem by using a mixture model of informative + uniform priors: an additional training set drawn from uniform prior distributions ($\sim$~30\% of the total set) is mixed into the original training set to ensure sufficient training set density in the full parameter space.

For the data set, we impose a signal-to-noise (S/N) cut of 3 in the F444W band, which results in $\sim$~300 galaxies, reaching a $5\sigma$ depth of 24.7 magnitude in F444W. We show the observed uncertainties as a function of magnitudes in F444W in Figure~\ref{fig:jwst}. A subsample, shown in purple, has spectroscopic redshifts collected from the NASA/IPAC Extragalactic Database, GLASS \citep{Treu2015}, and MUSE \citep{Richard2021}. We utilize this spectroscopic sample to evaluate the accuracy of our photometric analysis.

The various depths of the imaging create the two distinct lines seen in the magnitude panel. Since our S/N cut is based on F444W, the noise properties in other bands tend to be even more complex. The noise model that has been adopted throughout this Letter is shown as red shades. It is constructed using simulated JWST data \citep{Wang2023}. This model assumes a Gaussian uncertainty at a fixed magnitude and is chosen for simplicity, as well as for its well-defined behavior toward the faint end. Such a simple noise model is also a well-suited test for the robustness of \sbip. As a further illustration of the advantage of \sbip, we consider a ``wider'' noise approach, in which more data points are encapsulated. Specifically, we train two noisy SBI models, one of the same training size as \sbip\, and one of 5 times the training size. This is done to test the effect of training set density. The larger training set increases the training time to $\sim$~4 days. The results are shown in blue in Figure~\ref{fig:jwst}. Also shown in the same figure is the comparison between photometric redshifts inferred using NS and \sbip. We discuss the various implications from these results below.

\section{Discussion and Conclusions\label{sec:concl}}

We aim to expand the applicability of SBI by removing the constraint that the simulated data used for model training must possess characteristics identical to those of the observed data. This Letter presents \sbip, which uses MC techniques to approximate OOD measurement errors via standard SBI evaluations, and missing data by marginalizing over nearby data realizations in the training set. \sbip\ retains the extremely fast inference speed of $\sim$~1 sec for objects in the observational training set distribution, and additionally permits parameter inference outside of the trained noise and data at $\sim$~1 min per object. 

In this section, we discuss the key findings in this work. We begin with the performance across different algorithms assessed with the validation set, and then proceed to the JWST application. While there is general agreements between NS and \sbip\ in both cases, the complexities of fitting early JWST data necessitate a dedicated discussion. 

To start, NS and baseline SBI show remarkable agreement. This is consistent with similar tests performed in \citet{Hahn2022}. We note that in this work we additionally fit redshift as a free parameter, instead of keeping it fixed: this is a significantly more challenging test due to the multi-modality introduced by variable distance on cosmological scales. In fact, Figure~\ref{fig:baseline} suggests that baseline SBI performs notably better than classic NS at locating the correct mode---though importantly this is a situation where the data are drawn from the prior distributions. Deriving accurate photometric redshifts is generally of broad interest and a well known challenge. It is thus encouraging to see that baseline SBI performs well in this capacity.
However, we caution that this particular performance comparison uses simple summary statistics---in particular, the posterior median. Although conventionally adapted for calibrating photometric redshifts, they do not capture the complexity of the full PDF; for instance, a biased MCMC run can show better summary statistics than an unbiased grid search \citep{Speagle2016}. We defer a more comprehensive assessment to future studies.

The second major finding is that \sbip\ can attain accurate posterior distributions in the presence of OOD measurement errors and missing data, using only tools native to the SBI technique. The following paragraphs discuss each scenario in more detail.

First, in Section~\ref{subsec:res_mcnoise} we see that \sbip\ (OOD) performs comparably to NS; particularly, the outlier fraction and the bias increase to a level that is more similar to NS. The agreement here is mainly caused by \sbip\ picking up the (incorrect) modes that are favored by NS via the MC process. This not an undesirable behavior. As data become noisy, it is reasonable to have multi-modal distributions due to inherent challenges to inferring photometric redshifts. \sbip\ (missing data) agrees with NS for the same reason. In contrast, NS sometimes fails to find the global maximum and thus will report overly-confident incorrect answers.

Second, the mean offset between the posterior median and the truth seen in \sbip\ (OOD) is comparable to that of NS in the majority of cases, as shown in Figure~\ref{fig:hist}. The only minor difference is that the outliers in the NS results with respect to the photometry with in-distribution errors cause some outliers here as well. This is consistent with the finding above; that is, NS occasionally finds the previously missed modes once the measurement errors become OOD. In contrast, naively passing the OOD errors through the baseline SBI performs substantially worse as expected. Comparing to NS and \sbip, the NMAD when using baseline SBI on objects with OOD errors increases by a factor of $>3$, and the MBE increases by a factor of $>7$. 

The changes in the standard deviations from the redshift posteriors inferred from unperturbed photometry to those from noisy photometry, shown in Figure~\ref{fig:hist}(b), however, appear to suggest different behaviors between \sbip\ and NS when dealing with noisy observations. This is just another manifestation of the earlier finding of the mode-finding ability of \sbip. The majority of the ``excess'' population, i.e., where $\rm \delta_{\sigma,SBI} > 0.5$, have multi-modal posteriors. Part of these modes are the incorrect modes found by NS on the photometry with in-distribution uncertainties; that is, the MC process helps \sbip\ more widely explore the parameter space, in a way that produces similar results to NS. Another set of modes, found only by SBI, have significant posterior mass in them; however, NS has missed these modes. 

Third, the shift in the posterior median from the truth seen in \sbip\ (missing bands) also agrees with that of NS. The shift in the standard deviation exhibits a similar trend as in the noisy case. Upon examination, we find that this again occurs in objects with multi-modal posteriors. \sbip\ tends to find an additional likely solution that is not seen by NS. 

In addition to the NS-based sampling challenges discussed above, there is an additional complexity here. In some cases, \sbip\ can weight well-separated modes differently than NS. This is due to challenges in \sbip\ rather than NS. There are two sometimes-overlapping causes. The training set can by chance be sparsely sampled in the parameter space where a solution is, and hence it is difficult to find NNs that can produce this solution. This can be solved by more densely sampling parameter space in the training set. On the other hand, our priors can explicitly favor or disfavor a solution. This is a generic problem in SBI, as the training set must be generated following the prior density; the fact that the training set is also used to approximate missing bands effectively produces additional dependence on the accuracy of the model priors. The effect of Bayesian priors on parameter inference is a well-known challenge, and hence is not discussed further here.

Turning now to the test done on JWST observations, we find that a standard SBI model trained with more generous noise models shows degraded performance. For a noisy model trained using the same training set size as \sbip, the scatter, outlier fraction, and bias all show $>4$ changes, whereas for a second noisy model trained using five times the training set size, as shown in Figure~\ref{fig:jwst}, the statistics improves considerably, but the scatter more than doubles. This trend suggests that the training set density has to be ensured when considering a wider noise approach. The results may be further improved if constructing a noise model from the photometric level, the detailed analysis of which is out of the scope of this Letter.
\sbip, however, performs comparably to the traditional NS approach, as clearly indicated by the statistics shown in the same figure. This demonstrates that \sbip\ is highly effective for noise outliers. These results are particularly encouraging considering the fact that our simple Gaussian noise model does not account for the varying depths in the observations, i.e., most objects are not particularly close to mean noise in our toy noise model.
Taken together, the choice of whether to use \sbip\ to model a larger fraction of the total sample, or increase the width of the noise distribution in the training set to achieve an overall shorter runtime, but at the expense of increased model complexity and uncertainty, is best made on a case-by-case basis. 

Interestingly, the photometric redshifts vs. spectroscopic redshifts plot in Figure~\ref{fig:jwst} shows that, while in most cases \sbip\ and NS consistently find the correct solution, \sbip\ produces more realistic uncertainties. The latter is most noticeable when NS assigns small uncertainties to the wrong redshifts. In principle, the modes missed by NS can be avoided by substantial increases in the accuracy settings in NS and thus the number of models called, at the cost of increased runtime. Here, we do not attempt to completely remove this error, but instead adopt realistic \texttt{dynesty} settings already more strict than those typically used in the literature (e.g., \citealt{Mathews2023}), which keeps the time per fit below an hour using a neural net emulator. We note that this particular problem creates a more noticeable difference in tests done on the real data, likely because the observations have less well-behaved properties---multiple bands having OOD errors are common in this data set, and also we are still at the early stage of calibrating JWST/NIRCam.

Nevertheless, it is remarkable to see that \sbip\ is able to find additional likely modes / more realistic uncertainties in the posterior distribution. This has important implications for photometric redshift inference. Given the rich literature on photometric redshifts (see \citealt{Newman2022} for a recent review), here we only reiterate that it is a well-known challenge to assign correct photometric redshift uncertainties for a simple reason: the most important information in determining a photometric redshift comes from the position of spectral breaks, (e.g., the dropout technique introduced in \citealt{Steidel1996}), but a Balmer break can be confused with a Lyman break or strong emission lines (e.g., \citealt{Dunlop2007,McKinney2023,Zavala2023}). In other words, low-$z$ objects contaminate high-$z$ detections when their photometry breaks in similar locations, meaning that breaking this degeneracy is very challenging and can only be done with subtle spectral features or strong priors on galaxy evolution. In the context of MCMC/NS, this challenge translates to a difficulty in finding all the probable modes. These sampling methods are known to occasionally miss a solution, and thus produce underestimated uncertainties. Our results indicate that \sbip\ is a promising solution to this particular problem. The slightly worsened statistics of NMAD and MBE may be improved by a larger training set, and/or further optimization of the flow architecture.

To conclude, heteroskedastic data and uncertainties are often unavoidable in astronomical surveys due to missing observations and/or variable uncertainties determined by sky and telescope conditions. The now expanded applicability of SBI, permitted by \sbip, therefore has broad implications for the use of SBI in astronomical contexts. In the future, we intend to apply \sbip\ to broader surveys. High-dimensional SED modeling has thus far been limited to relatively small samples due to the high computational requirements. Large-scale applications of some of the most sophisticated models for galaxy evolution will establish new, in-depth censuses of cosmic history, and also permit a new depth to research in extragalactic astronomy.

\section*{acknowledgments}
We thank John Weaver and Kate Whitaker for providing us with the reduced JWST photometry prior to the public release.
B.W. is supported by the Institute for Gravitation and the Cosmos through the Eberly College of Science. This research received funding from the Pennsylvania State University's Institute for Computational and Data Sciences through the ICDS Seed Grant Program. Computations for this research were performed on the Pennsylvania State University's Institute for Computational and Data Sciences' Roar supercomputer. This publication made use of the NASA Astrophysical Data System for bibliographic information.

\,

\facilities{JWST (NIRCam)}
\software{Astropy \citep{2013A&A...558A..33A,2018AJ....156..123A,2022ApJ...935..167A}, Corner \citep{2016JOSS....1...24F}, Dynesty \citep{Speagle2020}, Matplotlib \citep{2007CSE.....9...90H}, NumPy \citep{2020Natur.585..357H}, Prospector \citep{Johnson2021}, PyTorch \citep{2019arXiv191201703P}, sbi \citep{tejero-cantero2020}, SciPy \citep{2020NatMe..17..261V}}

\bibliography{sbi_wang.bib}

\end{document}